\documentclass[12pt,thmsa]{article}
\usepackage{sw20aip}

\input{tcilatex}
\begin{document}

\author{Efrain J. Ferrer and Vivian de la Incera \\
State University of New York at Fredonia\\
SUNY-FRE-98-05\\
{\small \ Talk given at SILAFAE'98. April 8 -11, 1998. }\\
{\small San Juan, Puerto Rico}}
\title{Yukawa Interactions and Dynamical Generation of Mass in an External Magnetic
Field\thanks{%
This work has been supported in part by NSF grant PHY-9722059}}
\maketitle

\begin{abstract}
In this work we study the dynamical generation of a fermion mass induced by
a constant and uniform external magnetic field in an Abelian gauge model
with a Yukawa term. We show that the Yukawa coupling not only enhances the
dynamical generation of the mass, but it substantially decreases the
magnetic field required for the mass to be generated at temperatures
comparable to the electroweak critical temperature. These results indicate
that if large enough primordial magnetic fields were present during the
early universe evolution, the field-induced generation of fermion masses,
which in turn corresponds to the generation of fermion bound states, may
play an important role in the electroweak phase transition.
\end{abstract}

In this talk I would like to speak about a recently found phenomenon known
as the catalysis of chiral symmetry breaking due to a magnetic
(chromomagnetic, hypermagnetic) field and its possible implications for the
electroweak phase transition. The essence of the catalysis of chiral
symmetry breaking lies in the dimensional reduction in the dynamics of
fermion pairing in the presence of a magnetic field\cite{mirans}$.$ Due to
such a dimensional reduction, the magnetic field catalyses the generation of
a fermion condensate, and consequently, of a dynamical fermion mass, even in
the weakest attractive interaction between fermions.

An important aspect of the magnetic field induced generation of a dynamical
mass (MIGDM) is related to its possible cosmological consequences. The
existence in the early universe of a magnetic field induced dynamical mass,
as well as the fermion condensate associated with it, would require the
presence of very large primordial magnetic fields during the early stages of
the universe evolution. It is worth to note, however, that at present large
primordial magnetic fields in the early universe do not seem an impossible
option. As it is known, such large fields may be needed to explain the
large-scale galactic magnetic fields $\sim 10^{-6}$G observed in our own, as
well as in other galaxies. There are several primordial field generating
mechanisms which typically predict fields as large as 10$^{24}G$ during the
electroweak phase transitions. Moreover, Ambj\o rn and Olesen\cite{olesen}
have claimed that seed primordial fields even larger, $\sim 10^{33}G,$ would
be necessary at the electroweak scale to explain the observed galactic
fields.

It has been speculated\cite{mirans} that the character of electroweak phase
transition could be affected by the MIGDM. However, the results of Lee,
Leung and Ng\cite{lee}$,$ and of Gusynin and Shovkovy\cite{shoko} seem to
indicate that the fields required for the MIGDM to be important at the
electroweak scale are too large ($\sim 10^{42}G$) to be realistically
attainable. It must be pointed out however that their results were found
from the study of MIGDM in QED at finite temperature, and it is reasonable
to expect that their conclusions may change in the context of the
electroweak model when a richer set of interactions enters in scene.

In the present paper, we show that even in a simpler toy model, which
retains some of the attributes of the electroweak theory, the effect of new
couplings, like Yukawa couplings, can substantially change the order of
magnitude of the magnetic field required for the dynamical fermion mass to
be nonzero at the electroweak scale.

The model under study will be an Abelian gauge model of massless fermions
with Yukawa interaction in the presence of a constant magnetic field. By
solving the Schwinger -Dyson equation for the fermion propagator in the
ladder approximation, we prove that the Yukawa interaction enhances the
dynamical generation of a magnetic-field-induced fermion mass. We study the
same model at finite temperature, calculating the critical temperature at
which the field-induced fermion mass disappears. For a Yukawa coupling of
order of the top coupling, the field strength, required to obtain a critical
temperature comparable to the electroweak critical temperature, is decreased
in 10 orders of magnitude as compared to the corresponding field strength in
QED.

These results indicate that if large enough primordial magnetic (or
hypermagnetic) fields were present at the electroweak scale, the
field-induced generation of fermion masses, which in turn corresponds to the
generation of fermion bound states, may play an important role in the
electroweak phase transition. Our main conclusion is that the Yukawa
interactions enhance the dynamical generation of fermion bound states and
masses in the presence of external magnetic (or hypermagnetic) fields, and
therefore, it is worth to study this effect in the unbroken phase of the
electroweak system.

Let us consider an Abelian gauge model with a Yukawa interaction described
by the Lagrangian

\begin{equation}
L=\frac{1}{4}F^{\mu \nu }F_{\mu \nu }+i\overline{\psi }\gamma ^{\mu
}\partial _{\mu }\psi -e\overline{\psi }\gamma ^{\mu }\psi A_{\mu }-\frac{1}{%
2}\xi (\partial _{\mu }A^{\mu })^{2}+\frac{1}{2}\partial _{\mu }\phi
\partial ^{\mu }\phi -\frac{\lambda }{4}\phi ^{4}-\sqrt{2}\lambda _{y}\phi 
\overline{\psi }\psi  \label{e1}
\end{equation}

Note that this Lagrangian has a U(1) gauge symmetry and a fermion number
global symmetry, but it does not have chiral symmetry, so the appearance of
a dynamical mass in this model cannot be linked to chiral symmetry breaking.
This is fine since our ultimate goal is to get insight of a possible
magnetic field induced dynamical mass in the electroweak theory, where
anyway there is no chiral symmetry to break. We are interested in the study
of this theory in the presence of a constant and uniform external magnetic
field $H$. Our aim is to find nonperturbative solutions of the
Schwinger-Dyson equation for the fermion propagator to investigate the
dependence of the dynamically generated mass on the Yukawa coupling. We must
point out however that, even though the appearance of the dynamical mass can
be traced to the existence of a fermion-antifermion condensate, there is no
Goldstone field produced in this case, because there is no continuous
symmetry broken by the fermion condensate in this simple model. This would
not be the case in the electroweak model, on which, if a fermion condensate
were catalyzed by the magnetic field, it would give rise to a nonzero vev of
the scalar field and hence to a Higgs-like spontaneous gauge symmetry
breaking.

The Schwinger-Dyson equation derived from this theory takes the form

\[
\overline{G}^{-1}(x,y)=G^{-1}(x,y)-ie\int d^{4}ud^{4}w\gamma \overline{G}%
(x,u)\overline{D}(x-w)\overline{\Gamma }_{\psi \psi A}+ 
\]
\begin{equation}
+i\sqrt{2}\lambda _{y}\int d^{4}ud^{4}w\overline{G}(x,u)\overline{S}(x-w)%
\overline{\Gamma }_{\psi \psi \phi }  \label{e2}
\end{equation}
Here $G$ refers to fermions, $D$ to gauge bosons and $S$ to scalar bosons. $%
\overline{\Gamma }_{\psi \psi A}$ and $\overline{\Gamma }_{\psi \psi \phi }$
are three-fields vertex functions. The bar indicates full Green functions.

Assuming that the couplings are small, we can take eq.(\ref{e2}) in the
ladder approximation. In this approximation the vertices will be taken bare
(we assume no coupling is running), the fermion propagator is taken full,
and the gauge and scalar boson propagators are taken in the tree
approximation. The change to momentum coordinates can be done with the help
of Ritus\cite{ritus} $E_{p}$ functions. Then the SD equation becomes

\[
(2\pi )^{4}\delta _{kk^{\prime }}\delta (p_{0}-p_{0}^{\prime })\delta
(p_{2}-p_{2}^{\prime })\delta (p_{3}-p_{3}^{\prime })\widetilde{\tsum }_{A}(%
\overline{p})= 
\]

\[
=ie^{2}\int d^{4}xd^{4}x^{\prime }\sum_{k"}\int \frac{dp"_{0}dp"_{2}dp"_{3}}{%
(2\pi )^{4}}\{\overline{E}_{p}(x)\gamma ^{\mu }E_{p"}(x)\frac{1}{\gamma
\cdot \overline{p}"-\widetilde{\tsum }_{A}(\overline{p}")}\times 
\]
\[
\times \overline{E}_{p"}(x^{\prime })\gamma ^{\nu }E_{p^{\prime }}(x^{\prime
})D_{\mu \nu }(x-x^{\prime })\}-i2\lambda _{y}^{2}\int d^{4}xd^{4}x^{\prime
}\sum_{k"}\int \frac{dp"_{0}dp"_{2}dp"_{3}}{(2\pi )^{4}}\{\overline{E}%
_{p}(x)E_{p"}(x)\times 
\]
\begin{equation}
\times \frac{1}{\gamma \cdot \overline{p}"-\widetilde{\tsum }_{A}(\overline{p%
}")}\overline{E}_{p"}(x^{\prime })E_{p^{\prime }}(x^{\prime })S(x-x^{\prime
})\}  \label{e3}
\end{equation}
where $\widetilde{\tsum }_{A}(\overline{p})$ is the fermion mass operator.
The $E_{p}$ matrix is defined as 
\begin{eqnarray}
E_{p}(x) &=&\sum\limits_{\sigma
}[N(n)e^{i(p_{0}x^{0}+p_{2}x^{2}+p_{3}x^{3})}D_{n}(\rho )]diag(\delta
_{\sigma 1},\delta _{\sigma -1},\delta _{\sigma 1},\delta _{\sigma -1}) 
\nonumber \\
&=&\sum\limits_{\sigma
}N(n)e^{i(p_{0}x^{0}+p_{2}x^{2}+p_{3}x^{3})}D_{n}(\rho )\Delta (\sigma )
\label{e4}
\end{eqnarray}
with $D_{n}(\rho )$ being the parabolic cylinder functions\cite{handbook}
with argument $\rho =\sqrt{2\left| eH\right| }(x_{1}-\frac{p_{2}}{eH})$ and
positive integer index

\begin{equation}
n=n(k,\sigma )\equiv k+\frac{eH\sigma }{2\left| eH\right| }-\frac{1}{2}%
,\quad n=0,1,2,...;\qquad \sigma =-1,1  \label{e5}
\end{equation}

Using the properties of the $E_{p}$ functions, the expression (\ref{e3}) can
be reduced to 
\[
\delta _{kk^{\prime }}\widetilde{\tsum }_{A}(\overline{p})=ie^{2}2\left|
eH\right| \sum_{k"}\sum_{\{\sigma \}}\int \frac{d^{4}\widehat{q}}{(2\pi )^{4}%
}\{\frac{e^{isgn(eH)(n-n"+\widetilde{n}"-n^{\prime })\varphi }}{\sqrt{n!n"!%
\widetilde{n}"!n^{\prime }!}}e^{-\widehat{q}_{\bot }^{2}}J_{nn"}(\widehat{q}%
_{\bot })J_{\widetilde{n}"n^{\prime }}(\widehat{q}_{\bot })\frac{1}{\widehat{%
q}^{2}}\times 
\]
\[
\times \left( g_{\mu \nu }-\left( 1-\xi \right) \frac{\widehat{q}_{\mu }%
\widehat{q}_{\nu }}{\widehat{q}^{2}}\right) \Delta \gamma ^{\mu }\Delta "%
\frac{1}{\gamma \cdot \overline{p}"-\widetilde{\tsum }_{A}(\overline{p}")}%
\tilde{\Delta}"\gamma ^{\nu }\Delta ^{\prime }- 
\]
\[
i2\lambda _{y}^{2}(2\left| eH\right| )\sum_{k"}\sum_{\{\sigma \}}\int \frac{%
d^{4}\widehat{q}}{(2\pi )^{4}}\{\frac{e^{isgn(eH)(n-n"+\widetilde{n}%
"-n^{\prime })\varphi }}{\sqrt{n!n"!\widetilde{n}"!n^{\prime }!}}e^{-%
\widehat{q}_{\bot }^{2}}J_{nn"}(\widehat{q}_{\bot })J_{\widetilde{n}%
"n^{\prime }}(\widehat{q}_{\bot })\frac{1}{\widehat{q}^{2}}\times 
\]
\begin{equation}
\times \Delta \Delta "\frac{1}{\gamma \cdot \overline{p}"-\widetilde{\tsum }%
_{A}(\overline{p}")}\tilde{\Delta}"\Delta ^{\prime }  \label{e6}
\end{equation}
where

\begin{equation}
J_{n_{p}n_{r}}(\widehat{q}_{\perp })\equiv \sum\limits_{m=0}^{\min
(n_{p},n_{r})}\frac{n_{p}!n_{r}!}{m!(n_{p}-m)!(n_{r}-m)!}[isgn(eH)\widehat{q}%
_{\perp }]^{n_{p}+n_{r}-2m}  \label{e6-a}
\end{equation}

\begin{equation}
\widehat{q}_{\mu }\equiv \frac{q_{\mu }\sqrt{2\left| eH\right| }}{2eH}\
,\qquad \mu =0,1,2,3  \label{e6-b}
\end{equation}

\begin{equation}
\widehat{q}_{\perp }\equiv \sqrt{\widehat{q}_{1}^{2}+\widehat{q}_{2}^{2}}%
,\quad \varphi \equiv \arctan (\widehat{q}_{2}/\widehat{q}_{1})  \label{e6-c}
\end{equation}

\begin{equation}
\overline{p}"=(p_{0}-q_{0},0,-sgn(eH)\sqrt{2\left| eH\right| k"},p_{3}-q_{3})
\label{e6-d}
\end{equation}

To solve equation (\ref{e6}) we must use the structure of the mass operator.
Although in the presence of the external field the mass operator's structure
is quite rich(see ref. [8]), we can use a more simple structure, which is in
agreement with the solution of the Ward identity within the present
approximation\cite{vv}$.$ Thus, we consider

\begin{equation}
\widetilde{\tsum }_{A}(\overline{p})=Z_{_{\Vert }}\gamma \cdot \overline{p}%
_{_{\Vert }}+Z_{\bot }\gamma \cdot \overline{p}_{_{\bot }}+m(\overline{p})
\label{e8}
\end{equation}

Note the separation between parallel- and perpendicular-
to-the-magnetic-field variables.

Using the structure of $\widetilde{\tsum }_{A}(\overline{p})$ given in (\ref
{e8}), taking into account that the contributions of large $\widehat{q}%
_{\bot }$ in (\ref{e6}) are suppressed by the factor $e^{-\widehat{q}_{\bot
}^{2}}$, and considering the infrared region $\overline{p}^{2}<<\left|
eH\right| $ and in particular the lower Landau level contributions $%
\overline{p}_{_{\bot }}=0,$ equation (\ref{e6}) (taken in the Feynman gauge)
can be simplified to lead to the following two equations in Euclidean space

\begin{equation}
Z_{_{\Vert }}\gamma \cdot \overline{p}_{_{\Vert }}=-4\lambda _{y}^{2}(\left|
eH\right| )\int \frac{d^{4}\widehat{q}}{(2\pi )^{4}}\frac{e^{-\widehat{q}%
_{\bot }^{2}}(1+Z_{_{\Vert }})\gamma \cdot (\overline{p}_{_{\Vert
}}-q_{_{_{\Vert }}})}{\widehat{q}^{2}\left[ (1+Z_{_{\Vert }})^{2}(\overline{p%
}_{_{\Vert }}-q_{_{_{\Vert }}})^{2}+m^{2}(\overline{p}_{_{\Vert
}}-q_{_{_{\Vert }}})\right] }  \label{e9}
\end{equation}
\begin{equation}
m(\overline{p}_{_{\Vert }}-q_{_{_{\Vert }}})=4\left| eH\right|
(e^{2}+\lambda _{y}^{2})\int \frac{d^{4}\widehat{q}}{(2\pi )^{4}}\frac{e^{-%
\widehat{q}_{\bot }^{2}}m(\overline{p}_{_{\Vert }}-q_{_{_{\Vert }}})}{%
\widehat{q}^{2}\left[ (1+Z_{_{\Vert }})^{2}(\overline{p}_{_{\Vert
}}-q_{_{_{\Vert }}})^{2}+m^{2}(\overline{p}_{_{\Vert }}-q_{_{_{\Vert
}}})\right] }  \label{e10}
\end{equation}

The first equation has solution $Z_{_{\Vert }}=0$ if $\lambda
_{y}^{2}<<16\pi ^{2}.$ The second is the gap equation, which, in the
infrared limit that we are considering, has solution

\begin{equation}
m\simeq \sqrt{2\left| eH\right| }\exp \left[ -\sqrt{\frac{\pi }{\alpha +%
\frac{\lambda _{y}^{2}}{4\pi }}}\right]  \label{e11}
\end{equation}
$\alpha $ is the fine structure constant. The consistency of the
approximation requires $m<<\sqrt{\left| eH\right| },$ which is satisfied if $%
\alpha +\frac{\lambda _{y}^{2}}{4\pi }<<1,$ so the dynamical mass appears in
the weak coupling region of the theory. Note that because of the exponential
function in (\ref{e11}), small changes in the exponent can yield substantial
changes in the mass. For instance, for $\lambda _{y}\simeq 0.7$, a value
comparable to the top Yukawa coupling, the dynamical mass (\ref{e11}) is
five orders of magnitude larger than the mass found in QED\cite{mirans}$^{,}$%
\cite{lee}$,$ which is given by $m\simeq \sqrt{2\left| eH\right| }\exp
\left[ -\sqrt{\frac{\pi }{\alpha }}\right] .$

Finite temperature calculations can be done using the well known imaginary
time formalism. In that case the gap equation takes the form

\begin{equation}
m(\omega _{n^{\prime }},p)=(\alpha +\frac{\lambda _{y}^{2}}{4\pi })\frac{T}{%
\pi }\sum_{n=-\infty }^{\infty }\int\limits_{-\infty }^{\infty }\frac{%
dkm(\omega _{n},k)}{\omega _{n}^{2}+k^{2}+m^{2}(\omega _{n},k)}%
\int\limits_{0}^{\infty }\frac{dk\exp (-\frac{x}{2\left| eH\right| })}{%
(\omega _{n}-\omega _{n^{\prime }})^{2}+(k-p)^{2}+x}  \label{e12}
\end{equation}
with $\omega _{n}=(2n+1)\pi T.$ From this equation one can show that the
critical temperature at which the dynamical mass $m$ vanishes is

\begin{equation}
T\simeq m(T=0)\simeq \sqrt{2\left| eH\right| }\exp \left[ -\sqrt{\frac{\pi }{%
\alpha +\frac{\lambda _{y}^{2}}{4\pi }}}\right]  \label{e13}
\end{equation}

Therefore, for $\lambda _{y}\simeq 0.7,$ we can estimate the magnetic field
required to have a critical temperature (the temperature at which the
dynamical mass vanishes) comparable to the electroweak critical temperature.
Such a critical field is $H\approx 10^{32}G.$ That is, thanks to the Yukawa
interaction the critical field has decreased in 10 orders of magnitude as
compared to its corresponding value in QED ($\lambda _{y}\simeq 0).$

We conclude that the Yukawa interactions enhance the dynamical generation of
fermion bound states and masses in the presence of external magnetic (or
hypermagnetic) fields, and therefore, it is worth to study this effect in
the unbroken phase of the electroweak system.

\begin{description}
\item  
\begin{quote}
\textbf{Acknowledgments}
\end{quote}
\end{description}

It is a pleasure to thanks D. Caldi, C. N. Leung, Y. J. Ng and I. A.
Shovkovy for very useful discussions. Our special thanks to V. A. Miransky
for enlightening discussions and for calling our attention to the basic
papers of ref.[3].

\end{document}